\documentclass[11pt]{article}
\usepackage{setspace}
\begin{document}
\title{{\bf{\Large Voros product and noncommutative inspired black holes\footnote{This paper is based
on the work in references \cite{sunplb, sungrg, sunijmpa}. Some new material on noncommutative quantum mechanics in
three spatial dimensions (not discussed in these papers) has also been included. } 
 }}}

\author{
{\bf {\normalsize Sunandan Gangopadhyay}$^{a,b}
$\thanks{sunandan.gangopadhyay@gmail.com, sunandan@iucaa.ernet.in, sunandan@bose.res.in}}\\
$^{a}$ {\normalsize Department of Physics, West Bengal State University, Barasat, India}\\
$^{b}${\normalsize Visiting Associate in Inter University Centre for Astronomy $\&$ Astrophysics,}\\
{\normalsize Pune, India}\\[0.3cm]
}
\date{}

\maketitle

\begin{abstract}
\noindent We emphasize the importance of the Voros product in defining noncommutative inspired black holes. 
The computation of entropy for both the noncommutative inspired Schwarzschild and Reissner-Nordstr\"{o}m
black holes show that the area law holds upto order $\frac{1}{\sqrt{\theta}}e^{-M^2/\theta}$. 
The leading correction to the entropy (computed in the tunneling formalism) is shown to be logarithmic.
The Komar energy $E$ for these black holes is then
obtained and a deviation from the standard identity $E=2ST_H$ is found at the order $\sqrt{\theta}e^{-M^2/\theta}$.
This deviation leads to a nonvanishing Komar energy at the extremal point $T_{H}=0$ of these black holes.
The Smarr formula is finally worked out for the noncommutative Schwarzschild black hole. 
Similar features also exist for a deSitter--Schwarzschild geometry.

\end{abstract}


\noindent Noncommutative inspired black holes \cite{Smail, Smailrev} has gained considerable interest recently.
In this paper, we discuss some issues concerning them.  
The main point of interest is that there is no clear cut connection of this type of noncommutativity with 
standard notions of a noncommutative (NC) spacetime where point-wise multiplications are replaced by appropriate star multiplications. 
Interestingly, we observe that the Voros star product \cite{voros} plays an important role in
obtaining the mass and charge densities of a static, spherically symmetric, smeared, 
charged particle-like gravitational source. In doing so, we have also extended our earlier
discussion of the formulation of NC quantum mechanics in two spatial dimensions to three dimensional NC space.
We then proceed to derive quantum corrections to the semiclassical Hawking temperature and
entropy for these black holes by the tunneling mechanism by going beyond 
the standard semiclassical approximation \cite{ban}.
Finally, we examine the status of the relation between the Komar energy \cite{komar}, entropy and 
Hawking temperature ($E=2ST_H$) in the context of these black holes.

To address the first issue, one needs to take recourse to the formulation and interpretational aspects of NC quantum
mechanics \cite{gouba, sunprl}. In two spatial dimensions, we observe that the inner product of the coherent
states $|z, \bar{z})$ (used in the construction of the wave-function of a “free point particle”) 
can be calculated by using a deformed completeness relation (involving the Voros product) among the coherent states
\begin{eqnarray}
\int \frac{\theta dzd\bar{z}}{2\pi}~|z, \bar{z})\star(z, \bar{z}|=1_{q}
\label{eg6}
\end{eqnarray}
where the Voros star product between two functions 
$f(z, \bar{z})$ and $g(z, \bar{z})$ is defined as
\begin{eqnarray}
f(z, \bar{z})\star g(z, \bar{z})=f(z, \bar{z})
e^{\stackrel{\leftarrow}{\partial_{\bar{z}}}
\stackrel{\rightarrow}{\partial_z}} g(z, \bar{z})~.
\label{eg7}
\end{eqnarray}
The wave-function of the “free point particle” on the NC plane \cite{spal, gouba} is given by
\begin{eqnarray}
\psi_{\vec{p}}=(p|z, \bar{z})=\frac{1}{\sqrt{2\pi\hbar^{2}}}
e^{-\frac{\theta}{4\hbar^{2}}\bar{p}p}
e^{i\sqrt{\frac{\theta}{2\hbar^{2}}}(p\bar{z}+\bar{p}z)}
\quad;\quad p=p_{x}+ip_{y}~,~ z=\frac{1}{\sqrt{2\theta}}(x+iy)
\label{wavefunction}
\end{eqnarray}
where the momentum eigenstates are normalised such that
$(p'|p)=\delta(p'-p)$ and satisfy the completeness relation
\begin{eqnarray}
\int d^{2}p~|p)(p|=1_{q}~.
\label{mom_comp}
\end{eqnarray}
It turns out that the Voros product plays a vital role in providing a consistent probabilistic interpretation of this wave-function.
These observations and interpretations now allow us to write down the overlap of two coherent states 
$|\xi, \bar{\xi})$ and $|w, \bar{w})$ using the
completeness relation for the position eigenstates in eq.(\ref{eg6}) 
\begin{eqnarray}
(w, \bar{w}|\xi, \bar{\xi})=\int \frac{\theta dzd\bar{z}}{2\pi}
~(w, \bar{w}|z, \bar{z})\star(z, \bar{z}|\xi, \bar{\xi})~.
\label{overlap}
\end{eqnarray}
It is easy to check that 
\begin{eqnarray}
(w, \bar{w}|z, \bar{z})=\frac{1}{\theta}e^{-|\omega-z|^2}
\label{solution}
\end{eqnarray}
satisfies the above equation. A straightforward dimensional lift of this solution
from two to three space dimensions immediately motivates one to write down the
mass and charge densities of a static, spherically symmetric,
smeared, charged particle-like gravitational source in three space dimensions as \cite{sunplb} 
\begin{eqnarray}
\rho_{\theta}^{(M)}(r)&=&\frac{M}{(4\pi\theta)^{3/2}}
\exp\left(-\frac{r^2}{4\theta}\right)\nonumber\\
\rho_{\theta}^{(Q)}(r)&=&\frac{Q}{(4\pi\theta)^{3/2}}
\exp\left(-\frac{r^2}{4\theta}\right)~.
\label{massden}
\end{eqnarray}  
Interestingly, the formulation of NC quantum mechanics in two spatial dimensions can be generalized to three dimensional
NC space also. The Voros star product can be defined in this odd dimensional space by identifying the appropriate basis
in which the state has to be represented. In complete analogy with the two dimensional case, one can
introduce the position basis in three spatial dimensions through an expansion in momentum basis as \cite{sinha}
\begin{eqnarray}
|\vec{x})=\frac{1}{(2\pi)^{3/2}}\int d^{3}p~ e^{-\frac{\theta}{4\hbar^{2}}\vec{p}^2}
e^{-\frac{i}{\hbar}\vec{p}.\vec{x}}|p)
\label{basis3dim}
\end{eqnarray}  
which once again satisfy a deformed completeness relation
\begin{eqnarray}
\int d^{3}x~ |\vec{x})\star(\vec{x}|=1_{q}
\label{basis3compl}
\end{eqnarray} 
where the Voros star product in three spatial dimensions is given by
\begin{eqnarray}
\star=e^{\frac{i}{2}\theta(\epsilon_{ij}-i\delta_{ij})\stackrel{\leftarrow}{\partial_{i}}
\stackrel{\rightarrow}{\partial_j}}~.
\label{starnew}
\end{eqnarray}
The overlap of two position states $|\vec{x})$ and $|\vec{x}')$ (using the completeness relation (\ref{basis3compl})) read 
\begin{eqnarray}
(\vec{x}'|\vec{x})=\int d^{3}x''~(\vec{x}'|\vec{x}'')\star(\vec{x}''|\vec{x})
\label{overlap10}
\end{eqnarray}
which yields by a simple inspection
\begin{eqnarray}
(\vec{x}'|\vec{x})=\frac{1}{(2\pi\theta)^{3/2}}e^{-\vec{r}^2/(2\theta)}~;~\vec{r}=\vec{x}-\vec{x}'~.
\label{solution10}
\end{eqnarray}
The formalism of NC quantum mechanics in three spatial dimensions, therefore, gives a specific representation of the   
Dirac delta function in three dimensions since
\begin{eqnarray}
\lim_{\theta\rightarrow 0}\frac{1}{(2\pi\theta)^{3/2}}e^{-\vec{r}^2/(2\theta)}=\delta^{(3)}(|\vec{r}|)
\label{rep10}
\end{eqnarray}
which immediately leads to eq.(\ref{massden}). The above discussion of obtaining the overlap between two position
states in three spatial dimensions based on the formalism of NC quantum mechanics is a direct derivation of eq.(\ref{massden})
in contrast to the arguement presented in two dimensions and also clearly brings out the important
part played by the Voros product in defining the mass and charge densities of the NC inspired black holes.

\noindent Solution of Einstein's equations with the above mass density incorporated
in the energy-momentum tensor leads to the following 
NC inspired Schwarzschild black hole metric \cite{Smail},\cite{Smailrev}
\begin{eqnarray}
ds^2 = -\left(1-\frac{4M}{r\sqrt\pi}\gamma(\frac{3}{2},
\frac{r^2}{4\theta})\right)dt^2 + \left(1-\frac{4M}{r\sqrt\pi}
\gamma(\frac{3}{2},\frac{r^2}{4\theta})\right)^{-1} 
dr^2 + r^2(d\tilde\theta^2+\sin^2\tilde\theta d\phi^2)~. 
\label{1.04}
\end{eqnarray}
The event horizon of the black hole can be 
found by setting $g_{tt}(r_h)=0$ in eq.(\ref{1.04}), which yields
\begin{eqnarray}
r_h=\frac{4M}{\sqrt\pi}\gamma(\frac{3}{2},\frac{r^2_h}{4\theta}).
\label{1.05}
\end{eqnarray}
The large radius regime ($\frac{r_{h}^2}{4\theta}>>1$) allow
us to expand the incomplete gamma function to 
solve $r_h$ by iteration. Keeping upto next to leading order 
$\sqrt{\theta}e^{-{M^2}/{\theta}}$ leads to
\begin{eqnarray}
r_h \simeq 2M\left[1-\frac{2M}{\sqrt{\pi\theta}}
\left(1+\frac{\theta}{2M^{2}}\right)e^{{-M^2}/{\theta}}
\right]~.  
\label{1.06}
\end{eqnarray}
Now for a general stationary, static and spherically 
symmetric space time, the Hawking temperature ($T_H$) 
is related to the surface gravity ($\kappa$) 
by the following relation \cite{Hawking}
\begin{eqnarray}
T_H=\frac{\kappa}{2\pi} 
\label{1.061}
\end{eqnarray}
where the surface gravity of the black hole is given by
\begin{eqnarray}
\kappa = \frac{1}{2}\left[\frac{dg_{tt}}{dr}\right]_{r=r_{h}}.
\label{1.07}
\end{eqnarray}
Hence the Hawking temperature for the NC inspired Schwarzschild black hole upto order 
$\sqrt{\theta}e^{-{M^2}/{\theta}}$ is given by
\begin{eqnarray}
T_{H}=\frac{1}{8{\pi}M}
\left[1-\frac{4M^3}{{\theta}{\sqrt{\pi\theta}}}
\left(1-\frac{\theta}{2M^{2}}-\frac{\theta^2}{4M^{4}}\right)
{e^{-M^2/\theta}}\right]~.
\label{1.08}
\end{eqnarray}
The Bekenstein-Hawking entropy can now be calculated from the first law of black hole thermodynamics which reads  
\begin{eqnarray}
dS_{BH}=\frac{dM}{T_H}~.
\label{1.1}
\end{eqnarray}
Hence the Bekenstein-Hawking entropy in the next to 
leading order in $\theta$ is found to be 
\begin{eqnarray}
S_{BH}=\int{\frac{dM}{T_H}}=4\pi M^2-16\sqrt{\frac{\pi}{\theta}}M^3
\left(1+\frac{\theta}{M^2}\right)e^{-M^2/\theta}~.
\label{1.11}
\end{eqnarray}
To express the entropy in terms of the NC horizon area ($A_{\theta}$), eq.(\ref{1.06}) is used to get
\begin{eqnarray}
A_{\theta}&=& 4\pi r^2_h=16\pi M^2-64\sqrt{\frac{\pi}{\theta}}
M^3\left(1+\frac{\theta}{2M^2}\right)e^{-M^{2}/\theta}+\mathcal{O}(\theta^{3/2}e^{-M^{2}/\theta}).
\label{1.12}
\end{eqnarray}
Comparing equations (\ref{1.11}) and (\ref{1.12}), 
we find that at the leading order in $\theta$ (i.e. upto order $\frac{1}{\sqrt{\theta}}e^{-{M^2}/{\theta}}$), 
the NC black hole entropy satisfies the area law 
(in the regime $\frac{r^2_h}{4\theta}>>1$)
\begin{eqnarray}
S_{BH}=\frac{A_{\theta}}{4}~.
\label{1.13}
\end{eqnarray}
We now look for corrections to the semiclassical area law upto leading order in $\theta$. 
 
\noindent To do so, we first compute the corrected Hawking temperature $\tilde{T}_{H}$. For that we use the tunneling method by going beyond the semiclassical approximation \cite{ban}. Considering the massless scalar particle tunneling
under the background metric (\ref{1.04}),  the corrected Hawking temperature is given by
\begin{eqnarray}
\tilde{T}_{H}=T_{H}\left[1+\sum_{i}\frac{\tilde{\beta}_{i}\hbar^{i}}
{(Mr_{h})^{i}}\right]^{-1}~.
\label{corr_temp}
\end{eqnarray}
Application of the first law of black hole thermodynamics once again with this corrected Hawking temperature, gives the following expression for the corrected entropy/area law :
\begin{eqnarray}
S&=& \frac{A_{\theta}}{4\hbar}+2\pi\tilde{\beta}_{1}\ln A_{\theta} - \frac{64\pi^{2}\tilde{\beta}_{2}\hbar^2}{A_{\theta}}+\mathcal{O}(\sqrt{\theta}e^{-\frac{M^2}{\theta}})~\nonumber\\        &=& S_{BH}+2\pi\tilde{\beta}_{1}\ln S_{BH}-\frac{16\pi^{2}\tilde{\beta}_{2}\hbar}{S_{BH}}+\mathcal{O}(\sqrt{\theta}e^{-\frac{M^2}{\theta}})~.
\label{corr_entr}
\end{eqnarray}
We now move on to solve Einstein's equations with both the mass and charge densities incorporated
in the energy-momentum tensor. This leads to the following 
NC inspired Reissner-Nordstr\"{o}m (RN) black hole metric \cite{Smailrev}
\begin{eqnarray}
ds^2 = -f_{\theta}(r) dt^2 + f_{\theta}^{-1}(r)dr^2 + 
r^2(d\tilde\theta^2+\sin^2\tilde\theta d\phi^2) 
\label{1.044}
\end{eqnarray}
where 
\begin{eqnarray}
\label{metric_coeff10}
g_{tt}(r)&=&g^{rr}(r)=f_{\theta}(r)=1-\frac{4M}{r\sqrt\pi}\gamma\left( \frac{3}{2},\frac{r^2}{4\theta}\right)
 + \frac{Q^{2}}{\pi r^{2}}\left[ F(r)+\sqrt{\frac{2}{\theta}}r\gamma\left( \frac{3}{2},\frac{r^2}{4\theta}\right) \right] 
\\
F(r)&=&\gamma^{2}\left( \frac{1}{2},\frac{r^{2}}{4\theta}\right) -\frac{r}{\sqrt{2\theta}}
\gamma\left( \frac{1}{2},\frac{r^{2}}{2\theta}\right)\nonumber.
\end{eqnarray}
The event horizon of the black hole can be 
found by setting $g_{tt}(r_h)=0$ in (\ref{metric_coeff10}).
Once again in the large radius regime ($\frac{r_{h}^2}{4\theta}>>1$), 
we can expand the incomplete gamma function to solve $r_h$ by iteration. Keeping upto order 
$\sqrt{\theta}e^{-r^{2}_{0}/(4\theta)}$, we obtain
\begin{eqnarray}
r_h&\simeq& r_{0}\left[ 1-\frac{r_{0}}{2\sqrt{\pi\theta}(r_{0}-M)}\left( 2M-\frac{Q^{2}}{\sqrt{2\pi\theta}}\right)
e^{-r^{2}_{0}/(4\theta)} \right.\nonumber\\
&&\left.+ \frac{Q^2}{\sqrt{2}\pi r_0 (r_0 -M)}e^{-r^{2}_{0}/(4\theta)}
- \sqrt{\frac{\theta}{\pi}}\frac{1}{r_{0}^2 (r_0 -M)}
\left(2Mr_0 -M^2 - \frac{2Q^2}{\sqrt{\pi}}\right)e^{-r^{2}_{0}/(4\theta)} 
\right]\nonumber\\  
\label{1.067}
\end{eqnarray}
where
\begin{equation}
r_{0}=M+\sqrt{M^{2}-Q^{2}}
\label{horcomm10}
\end{equation}
is the horizon radius of the commutative RN black hole. Now using eq(s)(\ref{1.061}, \ref{1.07}),
we obtain the Hawking temperature for the NC inspired RN black hole 
(upto order $\sqrt{\theta}e^{-r^{2}_{0}/(4\theta)}$)
\begin{eqnarray}
T_{H}&\simeq&\frac{\hbar}{2\pi r^{3}_{0}}\left[ Mr_{0}-Q^2+\frac{Mr_{0}^2}{\sqrt{\pi\theta}(r_{0}-M)}\left( 3M-r_{0}
-\frac{Q^{2}}{\sqrt{2\pi\theta}}-\frac{(r_{0}-M)}{2\theta}r_{0}^2\right)e^{-r^{2}_{0}/(4\theta)}\right.\nonumber\\  
&&\left. \quad\quad~~~~~+\frac{Q^{2}r_{0}^{4}}{\sqrt{2}4\pi\theta^{2}}e^{-r^{2}_{0}/(4\theta)}
+\frac{Q^2}{\sqrt{2}\pi}e^{-r^{2}_{0}/(4\theta)}+2\sqrt{\frac{\theta}{\pi}}\left(M-\frac{3Q^2}{r_0}\right)
e^{-r^{2}_{0}/(4\theta)}\right]~. 
\label{1.10za}
\end{eqnarray}
We shall now write down the first law of black hole thermodynamics 
in the case of a charged black hole to calculate the Bekenstein-Hawking entropy. 
It reads \cite{sunijmpa}
\begin{eqnarray}
S=\int\frac{dM}{T_{H}}+\int Y dQ
\label{first_law100}
\end{eqnarray}
where 
\begin{eqnarray}
\label{exp100}
Y&=&-\frac{\Phi_{H}}{T_{H}}-\frac{\partial}{\partial Q}\int\frac{dM}{T_{H}}\\
\Phi_{H}&=&\frac{Q}{r_h}~. \nonumber
\end{eqnarray}
Using eq(s)(\ref{1.067}, \ref{1.10za}, \ref{exp100}), the Bekenstein-Hawking entropy upto order  $\sqrt{\theta}e^{-r^{2}_{0}/(4\theta)}$ is found to be 
\begin{eqnarray}
S&=&\pi r^{2}_{0}\left[ 1-\frac{r_{0}}{\sqrt{\pi\theta}(r_{0}-M)}
\left(2M-\frac{Q^{2}}{\sqrt{2\pi\theta}}\right) e^{-r^{2}_{0}/(4\theta)}\right.\nonumber\\
&&\left. +\sqrt{\frac{\theta}{\pi}}\frac{4M}{r_0 (r_0 -M)^2}\left\{8M-5r_0 - \frac{Q^2}{\sqrt{2\pi\theta}}
\left(2-\frac{r_0}{M}\right)\right\}e^{-r^{2}_{0}/(4\theta)}\right]. 
\label{1.110}
\end{eqnarray}
In order to express the entropy in terms of the NC horizon area ($A_{\theta}$), we use eq.(\ref{1.067}) to obtain
\begin{eqnarray}
A_{\theta}&=& 4\pi r^2_h=4\pi r^{2}_{0}\left[ 1-\frac{r_{0}}{\sqrt{\pi\theta}(r_{0}-M)}
\left( 2M-\frac{Q^{2}}{\sqrt{2\pi\theta}}\right)e^{-r^{2}_{0}/(4\theta)}\right.\nonumber\\
&&\left.+ \frac{\sqrt{2}Q^2}{\pi r_0 (r_0 -M)}e^{-r^{2}_{0}/(4\theta)}
- \sqrt{\frac{\theta}{\pi}}\frac{2}{r_{0}^2 (r_0 -M)}
\left(2Mr_0 -M^2 - \frac{2Q^2}{\sqrt{\pi}}\right)e^{-r^{2}_{0}/(4\theta)}  \right].  
\label{1.120}
\end{eqnarray}
Comparing eq(s)(\ref{1.110}, \ref{1.120}), we find that at the next to leading order in $\theta$, 
the NC black hole entropy satisfies the area law in the regime $r^2_h /(4\theta)>>1$
\begin{eqnarray}
S=S_{BH}=\frac{A_{\theta}}{4\hbar}~.
\label{1.130}
\end{eqnarray}
To investigate the corrections to the semiclassical area law upto next to leading order in $\theta$, 
we once again need to compute the corrected Hawking temperature $\tilde{T}_{H}$. In this case, it reads \cite{ban}
\begin{eqnarray}
\tilde{T}_{H}=T_{H}\left[1+\sum_{i}\frac{\tilde{\beta}_{i}\hbar^{i}}
{(Mr_{h}-Q^{2}/2)^{i}}\right]^{-1}~.
\label{corr_temp10}
\end{eqnarray}
Application of the first law of black hole thermodynamics once again with this corrected Hawking temperature leads to the following expression for the corrected entropy/area law :
\begin{eqnarray}
S&=& \frac{A_{\theta}}{4\hbar}+2\pi\tilde{\beta}_{1}\ln A_{\theta} 
+\mathcal{O}(\sqrt{\theta}e^{-r_{0}^2/(4\theta)})~\nonumber\\      
&=& S_{BH}+2\pi\tilde{\beta}_{1}\ln S_{BH}+\mathcal{O}(\sqrt{\theta}e^{-r_{0}^2/(4\theta)})
\label{corr_entr100}
\end{eqnarray}
where $A_{\theta}$ and $S_{BH}$ are defined in (\ref{1.120}) and (\ref{1.130}) respectively. 

\noindent Finally, we proceed to investigate the status of the relation between the Komar energy $E$,
entropy $S$ and Hawking temperature $T_H$
\begin{eqnarray}
E=2ST_{H}
\label{komar_ener}
\end{eqnarray}
in the case of these NC inspired black holes. The expression for the Komar
energy $E$ for the NC inspired Schwarzschild metric (\ref{1.04}) is given by \cite{sunplb}
\begin{eqnarray}
E=\frac{2M}{\sqrt\pi}\gamma\left(\frac{3}{2},\frac{r^2}{4\theta}\right)
-\frac{Mr^3}{2\theta\sqrt{\pi\theta}}e^{-r^2/(4\theta)}~.
\label{komarmass}
\end{eqnarray}
This expression allows one to identify $M$ as the mass of the black hole since $E=M$ in the limit $r\rightarrow\infty$. This identification plays an important role as we shall see below.

\noindent The above expression computed near the event horizon of the black hole\footnote{The Komar energy computed near the event horizon of the black hole plays an important role in obtaining the coefficient of the logarithmic
correction term (\ref{corr_entr}) in the entropy \cite{sunplb}.}
upto order $\sqrt{\theta}e^{-{M^2}/{\theta}}$ gives
\begin{eqnarray}
E &=& M\left[1-\frac{2M}{\sqrt{\pi\theta}}
\left(\frac{2M^{2}}{\theta}+1\right)e^{{-M^2}/{\theta}}-\frac{1}{M}\sqrt{\frac{\theta}{\pi}}e^{{-M^2}/{\theta}}\right].
\label{komar1c}
\end{eqnarray}
Finally, using eqs.(\ref{1.08}), (\ref{1.11}) and (\ref{komar1c}),       
we obtain
\begin{eqnarray}
E&=&2ST_{H} +2\sqrt{\frac{\theta}{\pi}}e^{{-M^2}/{\theta}}+ 
\mathcal{O}(\theta^{3/2}e^{-M^{2}/\theta})\nonumber\\
&=&2ST_{H} +2\sqrt{\frac{\theta}{\pi}}e^{{-S}/{(4\pi\theta)}}+ 
\mathcal{O}(\theta^{3/2}e^{-S/(4\pi\theta)})
\label{komar2a}
\end{eqnarray}
where in the second line we have used eq.(\ref{1.11}) to replace $M^2$
by $S/(4\pi)$ in the exponent. Interestingly, we observe that the
relation $E=2ST_H$ gets deformed upto order $\sqrt{\theta}e^{-{M^2}/{\theta}}$ which is consistent with 
the fact that the area law also gets modified at this order. The deformation gives a nonvanishing Komar energy at the extremal point $T_H =0$ of these black holes \cite{sungrg}. Also, we have once again managed to write down the deformed relation in terms of the Komar energy $E$, entropy $S$ and the Hawking temperature $T_H$. Similar features are also present for a de-Sitter Schwarzschild geometry \cite{dym}.
Eq.(\ref{komar2a}) can also be written with $M$ being expressed in terms of the black hole parameters
$S$ and $T_H$ using eq.(\ref{komar1c}) 
\begin{eqnarray}
M&=&2ST_{H} +\frac{1}{2\pi\sqrt{\pi\theta}}\left(S+\frac{S^2}{2\pi\theta}+6\pi\theta\right)e^{-S/(4\pi\theta)}
+\mathcal{O}(\theta^{3/2}e^{-S/(4\pi\theta)}).
\label{nc_smarr}
\end{eqnarray} 
We name eq.(\ref{nc_smarr}) as the {\it{Smarr formula}} \cite{smarr} {\it{for NC inspired Schwarzschild black hole}}
since $M$ has been identified earlier to be the mass of the black hole. 

\noindent The expression for the Komar energy $E$ for the NC inspired RN metric (\ref{1.044}) is given by \cite{sunijmpa}
\begin{eqnarray}
E&=&(r_{0}-M)\left[ 1-\frac{Mr_{0}}{\sqrt{\pi\theta}(r_{0}-M)}\left( 1+\frac{r^{2}_{0}}{2\theta}\right)e^{-r^{2}_{0}/(4\theta)}\right.\nonumber\\
&&\left.+\frac{Q^{2}r^{3}_{0}}{\sqrt{2}4\pi\theta^{2}(r_{0}-M)}
e^{-r^{2}_{0}/(4\theta)}+\frac{Q^{2}}{\sqrt{\pi\theta}(r_{0}-M)^{2}}\left( 2M-\frac{Q^{2}}{\sqrt{2\pi\theta}}\right)e^{-r^{2}_{0}/(4\theta)}\right.\nonumber\\
&&\left. -2\sqrt{\frac{\theta}{\pi}}\frac{1}{r_0 (r_0 -M)}\left(M-\frac{3Q^2}{r_0}\right)e^{-r^{2}_{0}/(4\theta)}\right]~. 
\label{komar1b1}
\end{eqnarray}
Now using eq(s)(\ref{1.10za}, \ref{1.110}, \ref{komar1b1}), we obtain upto order $\sqrt{\theta}e^{-r^{2}_{0}/(4\theta)}$
\begin{eqnarray}
E&=&2ST_H -\sqrt{\frac{\theta}{\pi}}\frac{4M}{r_0 (r_0 -M)}\left\{7M-4r_0 - \frac{Q^2}{\sqrt{2\pi\theta}}
\left(2-\frac{r_0}{M}\right)\right\}e^{-r^{2}_{0}/(4\theta)}\nonumber\\
&=&2ST_H -\sqrt{\frac{\theta}{\pi}}\frac{4M}{r_0 (r_0 -M)}\left\{7M-4r_0 - \frac{Q^2}{\sqrt{2\pi\theta}}
\left(2-\frac{r_0}{M}\right)\right\}e^{-S/(4\pi\theta)}
\label{smarrdef10}
\end{eqnarray}
where in the second line we have used eq.(\ref{1.110}) to replace $r_{0}^2$ by $S/\pi$ in the exponent.
The above relation is analogous to the deformed relation between Komar energy $E$, entropy $S$
and semiclassical Hawking temperature $T_H$ derived in case of the NC inspired Schwarzschild
black hole \cite{sungrg}. However, in this case the relation is not amongst $E$, $S$ and $T_H$ only but
also involves the mass $M$ and charge $Q$ of the black hole. In the limit $Q\rightarrow0$, the above result
reduces to eq.(\ref{komar2a}) \cite{sungrg}.

\section*{Acknowledgments} I would like to thank Inter University Centre for Astronomy $\&$ Astrophysics, Pune, India
for providing facilities.


\end{document}